\documentclass[a4paper,11pt]{article}

\usepackage{contribution}

\newcommand{\weblink}[2][]{%
    \ifthenelse{\equal{#1}{}}%
    {\textnormal{\url{#2}}}%
    {\textnormal{\href{#2}{#1}}}%
}

\newcommand{\acknowledgements}[1]{%
  \bigskip\bigskip
  \textsf{\textbf{\Large Acknowledgements}} \\[2ex]
  {#1}
  \bigskip
}

\def\beq{\begin{equation}}
\def\eeq#1{\label{#1}\end{equation}}
\def\eeqn{\end{equation}}

\def\beqa{\begin{eqnarray}}
\def\eeqa#1{\label{#1}\end{eqnarray}}
\def\eeqan{\end{eqnarray}}

\let\bar=\overbar

\def\etal{{\it et al.}}

\def\Dslash{\not{\hbox{\kern-4pt $D$}}}
\def\dslash{\not{\hbox{\kern-2pt $\del$}}}

\def\msb{{\bar{\ssstyle M \kern -1pt S}}}

\newcommand{\contribution}[7][]{%
  \clearpage
  \thispagestyle{plain}
  \ifthenelse{\equal{#1}{}}
  {\hypersetup{pdftitle={#2}}}
  {\hypersetup{pdftitle={#1}}}
  \hypersetup{pdfauthor={{#3} {#4}}}
  {\centering\normalfont\LARGE\bfseries\sffamily #2 \par\nobreak}
  \lhead{}
  \chead{%
    \textit{\footnotesize XIV International Conference on Hadron Spectroscopy
      (\weblink[\textit{hadron2011}]{http://www.hadron2011.de}), 13-17 June 2011, Munich, Germany}%
  }
  \rhead{}
  \bigskip
  \begin{center}
    {#3} {#4}\ifthenelse{\equal{#6}{}}{}{\footnote{\weblink[#6]{mailto:#6}}}
    \ifthenelse{\equal{#7}{}}{}{#7} \\
    \textit{#5}
  \end{center}
  \bigskip
}

\renewcommand{\abstract}[1]{%
  \begin{center}
    \begin{minipage}{0.85\textwidth}
      \begin{footnotesize}
        #1
      \end{footnotesize}
    \end{minipage}
  \end{center}
  \bigskip
}

\begin{document}

%
%
%
%
%
{  

\makeatletter
\@ifundefined{c@affiliation}%
{\newcounter{affiliation}}{}%
\makeatother
\newcommand{\affiliation}[2][]{\setcounter{affiliation}{#2}%
  \ensuremath{{^{\alph{affiliation}}}\text{#1}}}
%

\contribution[Test of OZI violation with COMPASS]  
{Test of OZI violation in vector meson\\production with COMPASS}  
{Johannes}{Bernhard}  
{\affiliation[Institut f\"ur Kernphysik, Johannes-Gutenberg-Universit\"at Mainz]{1} \\
 \affiliation[CERN, Geneva]{2}}
{johannes.bernhard@cern.ch}
{\!\!$^,\affiliation{1}$ and Karin Sch\"onning \affiliation{2} on behalf of the COMPASS Collaboration}
%


\abstract{%
\vspace{- 0.5cm}
The COMPASS experiment at CERN SPS completed its data taking with hadron beams ($p$, $\pi$, $K$) in the years 2008 and 2009 by collecting a large set of data using different targets (H$_2$, Pb, Ni, W). These data are dedicated to hadron spectroscopy, where the focus is directed to the search for exotic bound states of quarks and gluons (\textit{hybrids, glueballs}). The production of such states is known to be favoured in glue-rich environments, $e.g.$ so-called OZI-forbidden processes. The OZI rule postulates that processes with disconnected quark line diagrams are forbidden. On the one hand, the study of the degree of OZI violation in vector meson production yields the possibility to learn more about the involved production mechanisms. On the other hand it helps to understand the nucleon's structure itself. Contrary to former experiments, the large data sample allows for detailed studies in respect to Feynman's variable $x_F$. We present results from the ongoing analysis on the comparison of $\omega$ and $\phi$ vector mesons production in $p\,p\,\longrightarrow\,p\,(\omega/\phi)\,p$, where the possibility of measuring the spin alignment of both vector mesons at the same time makes COMPASS unique.
}
%

\vspace{- 1cm}
\section{Introduction}
\vspace{- 0.3cm}
The OZI rule\,\cite{ref:OZI} declares that processes with disconnected quark line diagrams are forbidden. Though being phenomenological when first formulated, it could later be explained by QCD. It has been helpful in explaining a multitude of phenomena, for example the large branching fraction of $\phi$ decays into $K\overline{K}$ final states and the suppressed production of mesons with an $s\overline{s}$ component. The production of $\phi$ mesons in reactions with only non-strange hadrons in the initial state is, according to OZI, only allowed due to the deviation $\delta_V = 3.7^{o}$ from ideal mixing of $\omega$ and $\phi$. This gives a prediction of the cross section ratio $R =\sigma(AB\rightarrow\phi X)/\sigma(AB\rightarrow\omega X)$ of $4.2\,\cdot\,10^{-3}$\,\cite{ref:lipkin}, where $A$, $B$ and $X$ are non-strange hadrons. Though based on a very simplified picture, the predicted value is in most reactions surprisingly well fulfilled\,\cite{ref:sapozhnikov}. However, violations have been observed in $p\overline{p}$ annihilations at rest and $NN$ collisions as well as $NN$ and $N\pi$ collisions near the kinematic threshold\,\cite{ref:sibirtsev}. These violations have been interpreted as i) intermediate gluonic states\,\cite{ref:lindenbaum}, ii) a polarised strangeness component in the nucleon\,\cite{ref:ellis} or, near the kinematic threshold, iii) differences in the production mechanism and the nature of the meson-nucleon interaction.
The COMPASS experiment is well suited for precise tests of the OZI rule and can provide data in a new energy range. In the following, the COMPASS experiment and the recent results of the ongoing analysis will be presented.



COMPASS is a two-stage magnetic spectrometer\,\cite{ref:spectro} at the CERN SPS dedicated to structure studies and spectroscopy. It features large angular acceptance over a wide momentum range and is capable of particle identification with the means of a RICH detector. Neutral particles are detected with electromagnetic (ECAL1+2) and hadronic (HCAL1+2) calorimeters in both stages of the spectrometer. To confine data taking to the relevant processes of diffractive scattering and central production, a scintillator barrel detector (RPD) was surrounding the target and used to measure and trigger on recoiling protons.
\vspace{- 0.3cm}
\section{Analysis and discussion}
\vspace{- 0.3cm}
\begin{figure}[htb]
  \begin{center}
   \vspace{- 0.7cm}
   \includegraphics[width=0.47\textwidth]{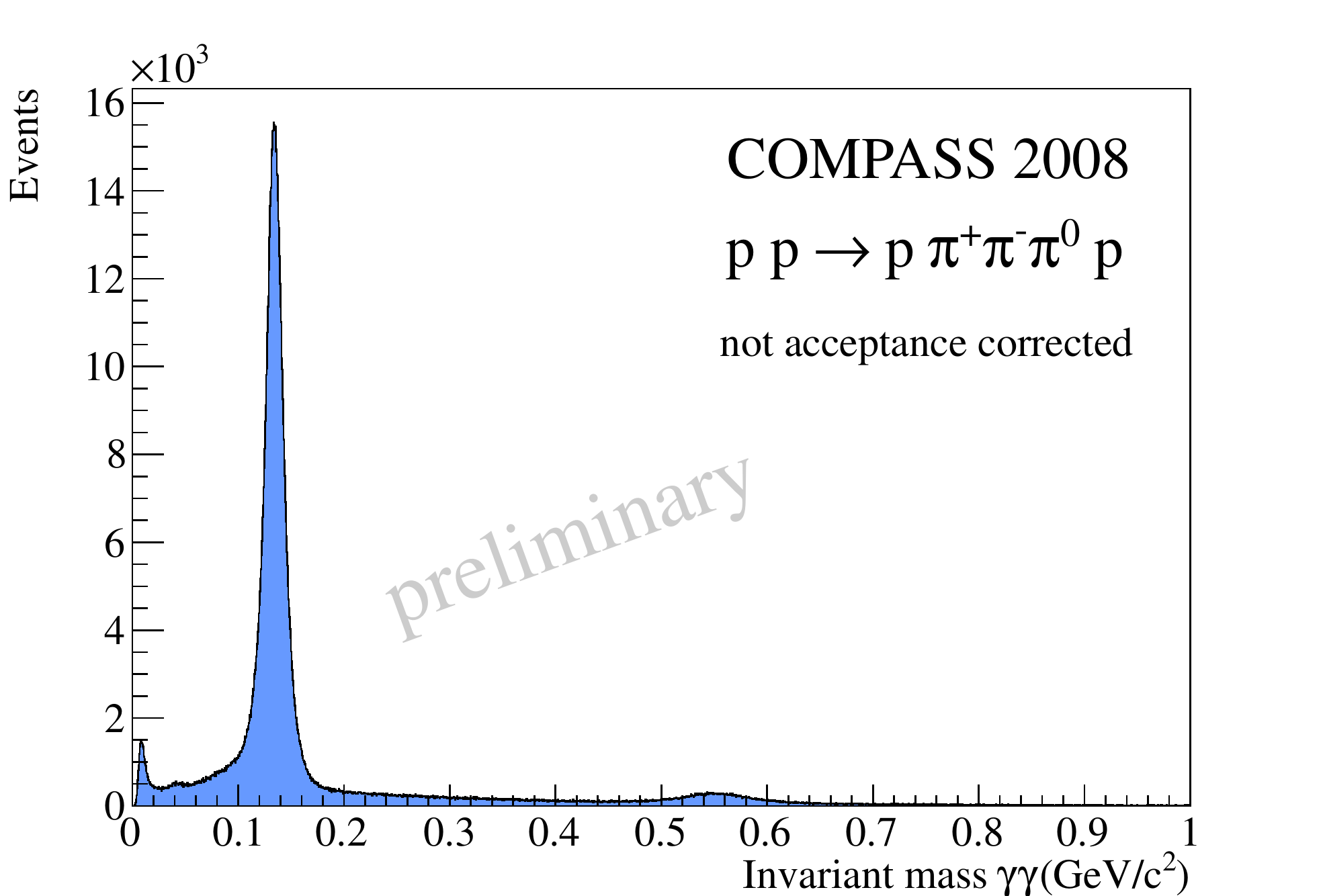}\includegraphics[width=0.4\textwidth]{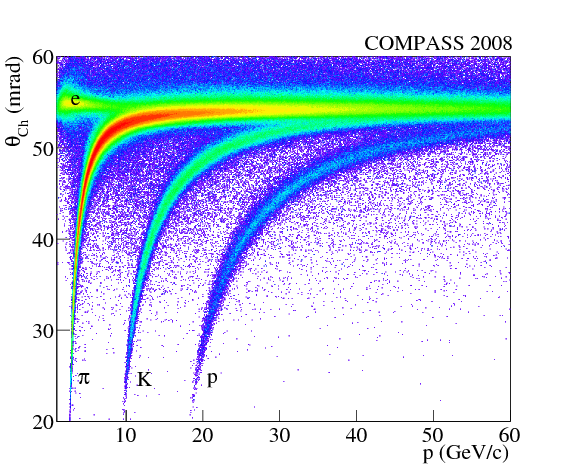}
   \caption{Left: Invariant mass distribution of $\gamma\gamma$ before cuts in the $p\,p\,\rightarrow\,p\,\pi^0\,\pi^+\,\pi^-\,p$ channel. The structures below the $\pi^0$ mass peak are artefacts of low energetic photon reconstruction due to secondary interactions in the detector material and to cuts in the reconstruction algorithm. They should not be mistaken for any physical signal. Right: Cherenkov angle (RICH detector) vs. momentum of charged particles.}
   \label{fig:ggrich}
 \end{center}
\end{figure}

\begin{figure}[htb]
\begin{center}
\includegraphics[width=0.45\textwidth]{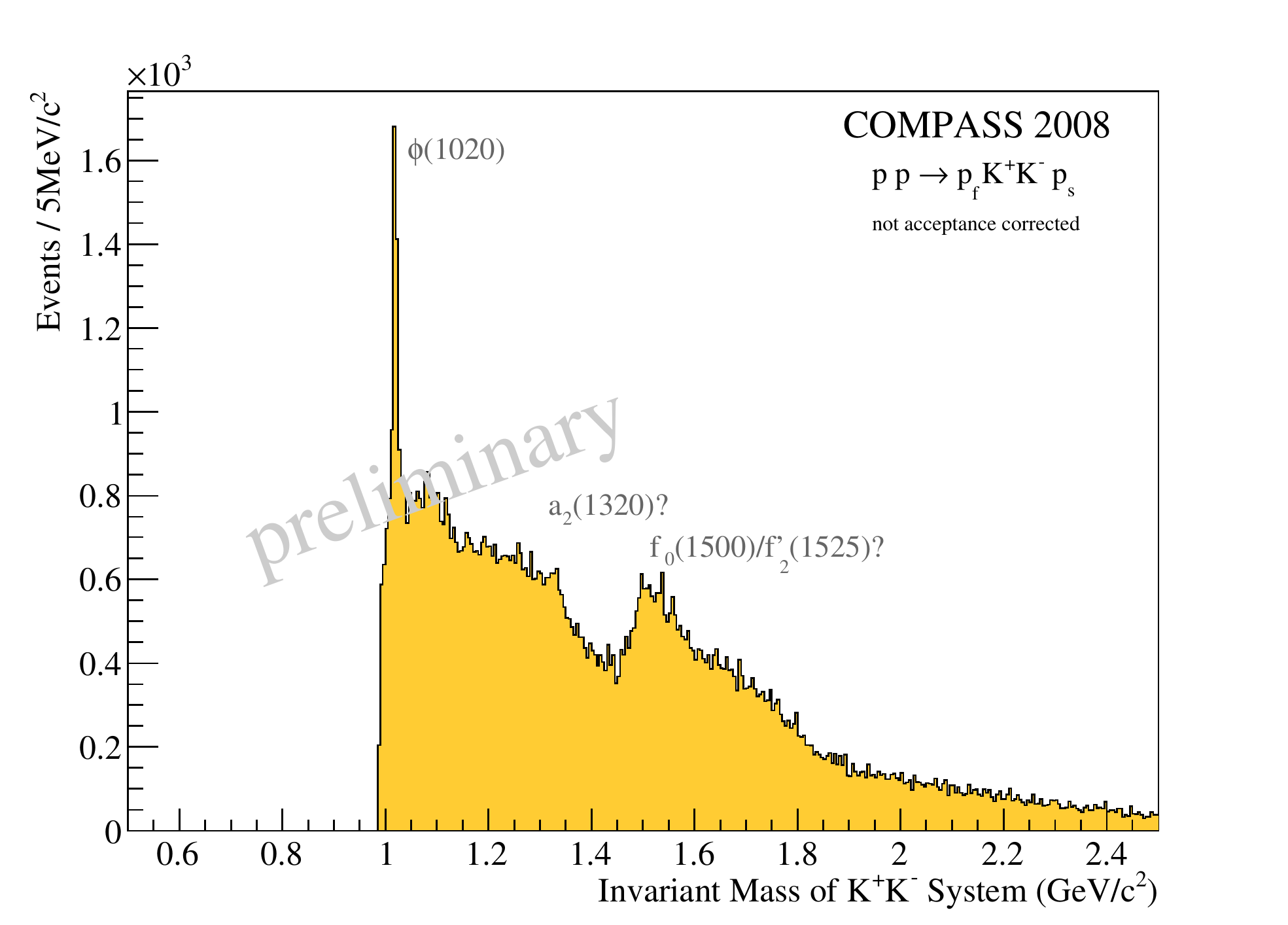}\includegraphics[width=0.45\textwidth]{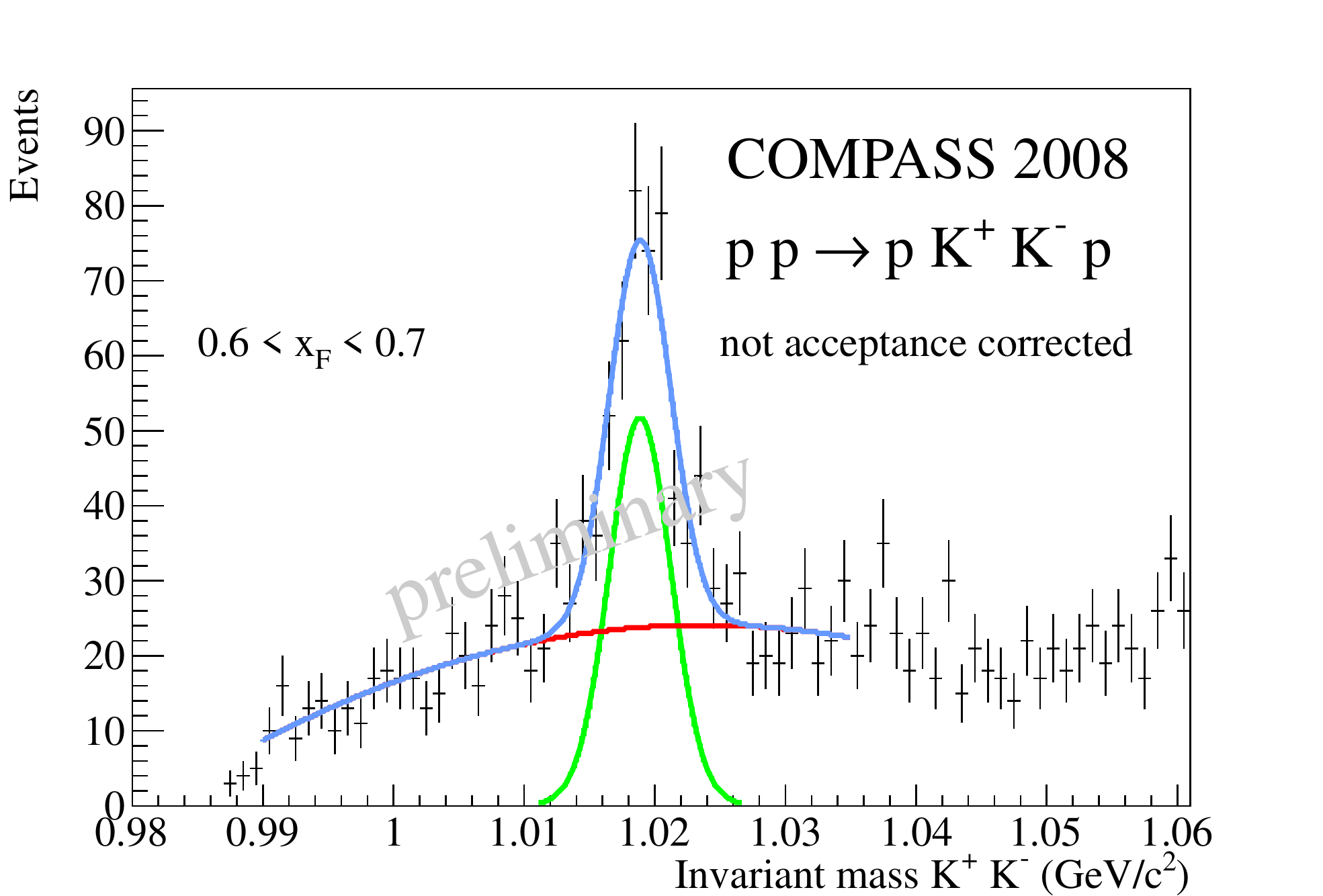}\\ 
\includegraphics[width=0.45\textwidth]{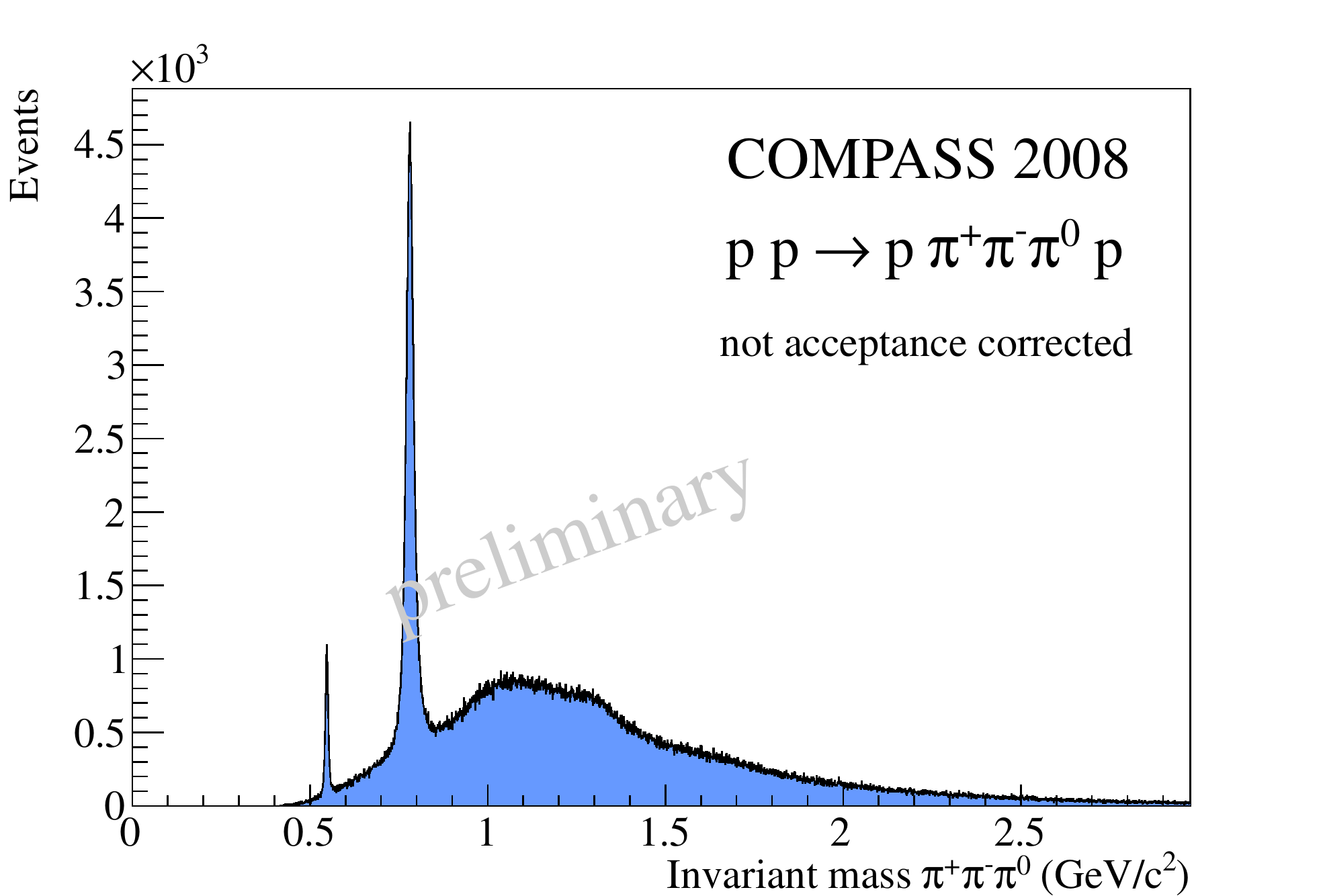}\includegraphics[width=0.45\textwidth]{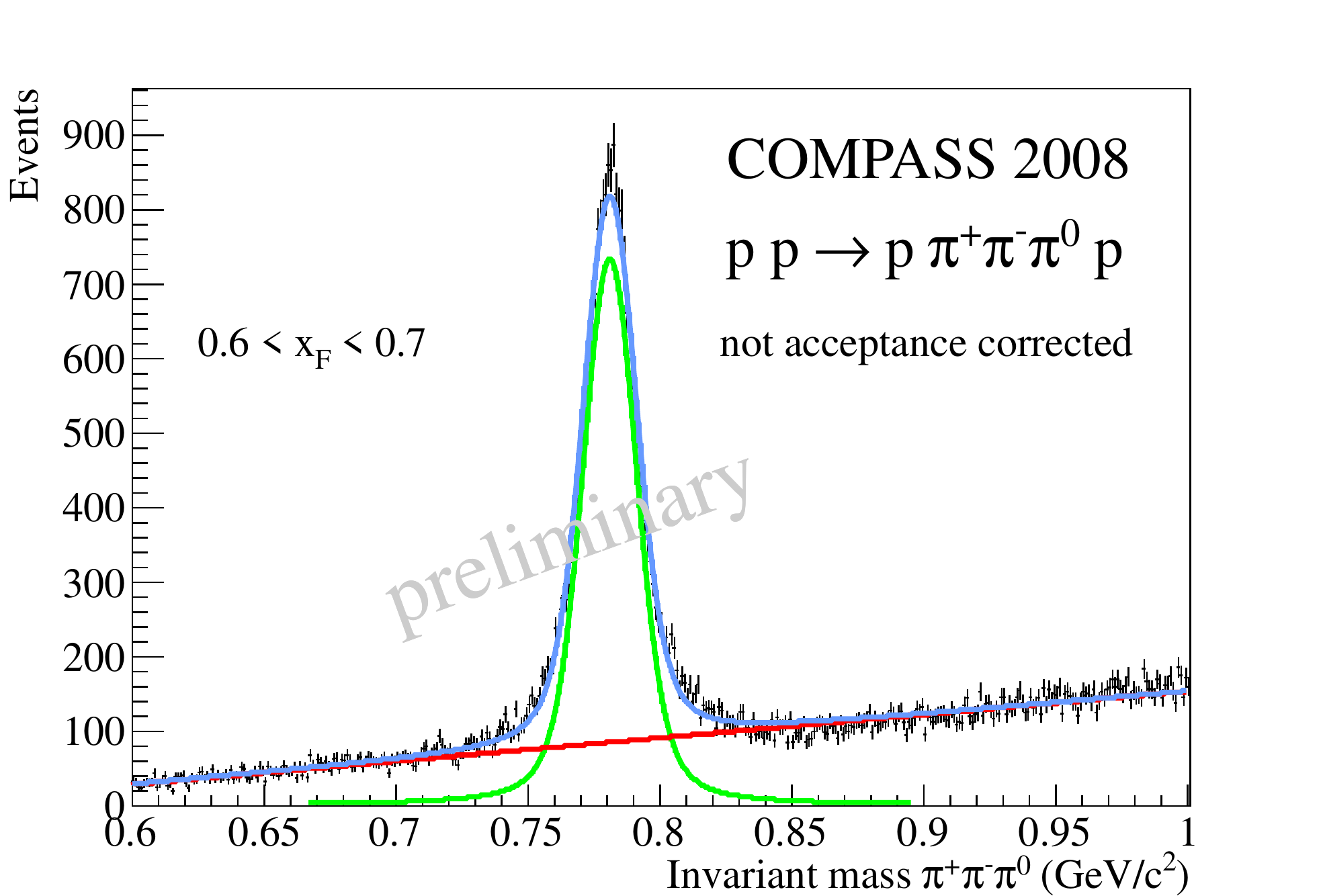}
\end{center}
\caption{Invariant mass distributions of the $K^+\,K^-$ and $\pi^0\,\pi^+\,\pi^-$ subsystems. Left: Full data sample. Right: $x_F$ of forward proton limited to $(0.6 < x_F < 0.7)$, with fitted curves as explained in the text.}
\label{fig:bins}
\end{figure}
The analysis presented here concentrates on the 2008 $p\,p$ data set, \textit{i.e.}\;one week of data of $190$\,GeV proton beams impinging on a liquid hydrogen target.
The two channels $p\,p\,\rightarrow\,p\,\omega\,p, \omega\,\rightarrow\,\pi^+\,\pi^-\,\pi^0$ and $p\,p\,\rightarrow\,p\,\phi\,p, \phi\,\rightarrow\,K^+\,K^-$ were compared. For both channels, events were selected with one tagged incoming proton, one recoil proton in the RPD and three outgoing charged tracks from the primary vertex. Furthermore, in the $\omega$ case, a $\pi^0$ candidate reconstructed from two photons (see the left panel of Fig.\,\ref{fig:ggrich}) and a positive RICH identification of the $\pi^+$ were required. In the $\phi$ case, positive RICH identification of the $K^+$ was required (for the RICH separation, see the right panel of Fig.\,\ref{fig:ggrich}).

For both decay channels, the total momentum of the final state was required to be within $\pm\,6$\,GeV/$c$ of the beam momentum and in addition, the recoiling proton and the forward $p\,\phi/p\,\omega$ system must be coplanar within $\pm\,2\,\sigma$ of the RPD's experimental resolution, \textit{i.e.} $0.28$\,rad. The data are binned in terms of $x_F$ of the forward proton. Fig.\,\ref{fig:bins} shows the invariant mass distributions of both the $K^+\,K^-$ and $\pi^0\,\pi^+\,\pi^-$ subsystems, respectively, for the total data sample presented here (upper left and lower left panel) and for the 
\begin{wrapfigure}[15]{l}{0.5\textwidth}
\begin{center}\vspace{- 0.7cm}
\includegraphics[width=0.5\textwidth]{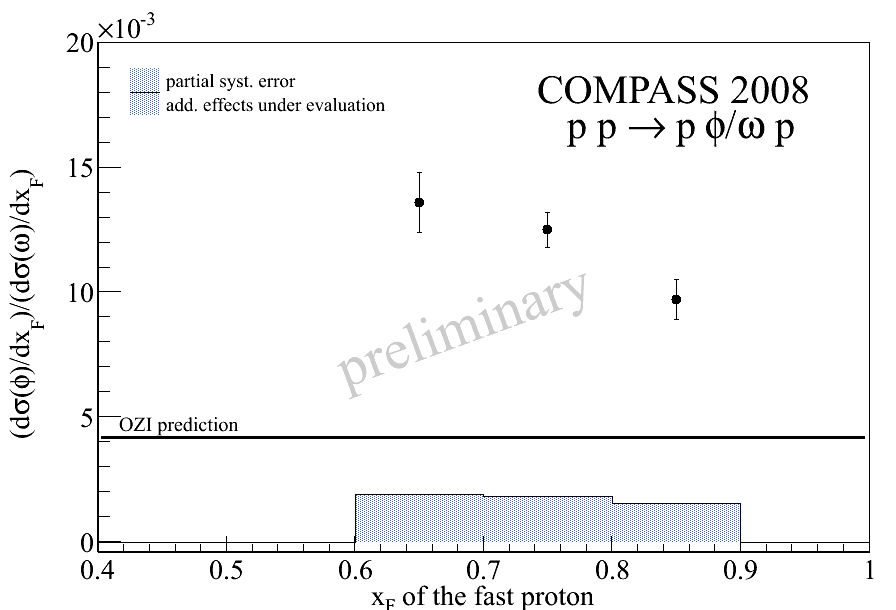}
\end{center}
\caption{Acceptance corrected ratio $R$ of $\phi$ and $\omega$ yields as a function of $x_F$ of the forward proton.}
\label{fig:result}
\end{wrapfigure}
subinterval $0.6 < x_F < 0.7$. A Breit-Wigner curve, convoluted with a Gaussian on top of a polynomial background was fitted to the data and the background was subtracted in order to obtain the yield of each vector meson. The yields were corrected for the branching ratio (89.2\% for $\omega\,\rightarrow\,\pi^+\,\pi^-\,\pi^0$ and 48.9\% for $\phi\,\rightarrow\,K^+\,K^-$) and acceptance, which was obtained by MC simulations. There is a systematic uncertainty in the photon reconstruction efficiency of 10\%, obtained by comparing the acceptance corrected yields of $\omega \rightarrow \pi^+\pi^-\pi^0$ and $\omega \rightarrow \gamma\pi^0$. 
The preliminary results are presented in Fig.\,\ref{fig:result}. The systematics shown include the photon reconstruction efficiency and the error of the BW fit, the latter being smaller than 1.5\%. Additional sources of systematic uncertainties like the RICH efficiency and the model dependence of the acceptance may also contribute non-negligibly but need further study and are therefore not quoted here. The COMPASS data show an OZI violation of approximately a factor of three.

\vspace{- 0.5cm}
\section{Outlook}
\vspace{- 0.3cm}
The 2009 proton data set will increase the statistics by a factor of $\approx$ 10 and will allow for finer binning. Furthermore, a dedicated comparative study of the vector meson spin alignment with respect to different quantisation axes, \textit{e.g.}\,as performed in\,\cite{ref:karin08} or suggested in\,\cite{ref:quing} is planned. This will give deeper insight into the underlying physics of diffractive and central processes at COMPASS energies. In particular, the studies of central systems will benefit from more knowledge of the production processes\,\cite{ref:hadron09}, but also the search for exotic states in multi-particle final states will profit due to better understanding of backgrounds, \textit{e.g.} in $\pi^-\,p\,\rightarrow\,K_s\,K^\pm\,\pi^\mp\,\pi^\mp\,p$\,\cite{ref:KKpipi}. 

\vspace{- 0.5cm}
\acknowledgements{This work was supported by the Bundesministerium f\"ur Bildung und Forschung (Germany).}

\vspace{- 0.7cm}

%

}  


\end{document}